# GPU-based fast Monte Carlo simulation for radiotherapy dose calculation

**Xun Jia, Xuejun Gu, Yan Jiang Graves, Michael Folkerts, and Steve B. Jiang**

Center for Advanced Radiotherapy Technologies and Department of Radiation Oncology, University of California San Diego, La Jolla, CA 92037-0843, USA

E-mail: sbjiang@ucsd.edu

Monte Carlo (MC) simulation is commonly considered to be the most accurate dose calculation method in radiotherapy. However, its efficiency still requires improvement for many routine clinical applications. In this paper, we present our recent progress towards the development a GPU-based MC dose calculation package, gDPM v2.0. It utilizes the parallel computation ability of a GPU to achieve high efficiency, while maintaining the same particle transport physics as in the original DPM code and hence the same level of simulation accuracy. In GPU computing, divergence of execution paths between threads can considerably reduce the efficiency. Since photons and electrons undergo different physics and hence attain different execution paths, we use a simulation scheme where photon transport and electron transport are separated to partially relieve the thread divergence issue. High performance random number generator and hardware linear interpolation are also utilized. We have also developed various components to handle fluence map and linac geometry, so that gDPM can be used to compute dose distributions for realistic IMRT or VMAT treatment plans. Our gDPM package is tested for its accuracy and efficiency in both phantoms and realistic patient cases. In all cases, the average relative uncertainties are less than 1%. A statistical t-test is performed and the dose difference between the CPU and the GPU results is found not statistically significant in over 96% of the high dose region and over 97% of the entire region. Speed up factors of 69.1 ~ 87.2 have been observed using an NVIDIA Tesla C2050 GPU card against a 2.27GHz Intel Xeon CPU processor. For realistic IMRT and VMAT plans, MC dose calculation can be completed with less than 1% standard deviation in 36.1~39.6 sec using gDPM.

**1. Introduction**

Radiation dose calculation plays a crucial role in radiotherapy treatment planning and verification. Monte Carlo (MC) simulation is known to be the most accurate dose calculation method for radiotherapy. In this method, one computes how a particle propagates step by step according to fundamental physics principles. To achieve a desired level of statistical accuracy, a large number of particle histories are needed, consuming a long computation time. During past a few decades, a number of MC algorithms have been developed, such as EGS4/5 (Nelson *et al.*, 1985; Bielajew *et al.*, 1994; Hirayama *et al.*, 2010), EGSnrc (Kawrakow, 2000), MCNP (Briesmeister, 1993), PENELOPE (Baró *et al.*, 1995; Salvat *et al.*, 1996; Sempau *et al.*, 1997; Salvat *et al.*, 2009), GEANT4 (Agostinelli *et al.*, 2003). Meanwhile, efforts have been devoted to develop high efficiency dose calculation packages for radiotherapy, for instance VMC++ (Kawrakow *et al.*, 1996), MCDOSE/MCSIM (Ma *et al.*, 1999; Li *et al.*, 2000; Ma *et al.*, 2002), and DPM (Sempau *et al.*, 2000), etc. These packages employ variance reduction techniques or simplify particle transport physics to gain computation speed. Despite these developments along with the increasing of CPU clock speed, there is still room for improving the efficiency of MC dose calculation.

Porting MC packages onto parallel computing architectures is a straightforward way for boosting their efficiency. Since each computing unit can work on a portion of the total particle histories without interfering with each other, roughly linear scalability of computation efficiency with respect to the number of computing units is commonly expected. This is indeed the case for CPU cluster based MC simulation. For instance, DPM has been parallelized on a CPU cluster and almost linear speed-up has been observed with the number of processors, when up to 32 nodes are used (Tyagi *et al.*, 2004). Recently, graphics processing unit (GPU) has been utilized to speed up computationally intensive tasks in medical physics. Though GPU has been demonstrated to be extremely powerful in solving many problems (Xu and Mueller, 2005; Sharp *et al.*, 2007; Yan *et al.*, 2008; Samant *et al.*, 2008; Jacques *et al.*, 2008; Hissoiny *et al.*, 2009; Men *et al.*, 2009; Gu *et al.*, 2009; Jia *et al.*, 2010b; Gu *et al.*, 2010; Men *et al.*, 2010a; Men *et al.*, 2010b; Gu *et al.*, 2011), it is quite hard to achieve high speed-up factors for MC dose calculations on GPU for the following two inherent conflicts between the stochastic nature of a MC process and the GPU hardware architecture. First, GPU employs an architecture called single-instruction multiple-thread (SIMT) (NVIDIA, 2010b), where the multiprocessor of a GPU executes a program in groups of 32 parallel threads termed *warps*. If the paths for threads within a warp diverge due to, *e.g.*, some *if-else* statements, the warp serially executes one thread at a time, while putting all other threads in an idle state. Thus, high computation efficiency is only achieved when all threads in a warp process together along a same execution path. Unfortunately, in a MC calculation the work paths on different threads are statistically independent, essentially resulting in an almost serialized execution within a warp. Second, GPU memory speed is typically very slow compared to CPU memory and it is quite expensive to have frequent and random memory access. In the MC simulation, all threads share the usage of GPU global memory and each thread visits different memory addresses in a completely unpredictable pattern. This fact also slows down the MC simulation considerably.

We have previously developed a GPU-based MC dose calculation package, gDPM v1.0 (Jia *et al.*, 2010a), based on a publically available dose calculation package, DPM. By simply distributing particles to all GPU threads and treating them as if they were independent computational units, it is found that only 5.0~6.6 times speedup can be achieved due to the aforementioned intrinsic conflicts between the randomness of the MC simulation and the GPU SIMT architecture. Recently, Hissoiny *et. al.* (Hissoiny *et al.*, 2011) have developed a MC dose calculation package, GPUMCD. By integrating various particle transport physics reported in literature and rewrite the code specifically designed for GPU, high computation efficiency has been reported. Despite its great speed, this code does not contain necessary features for dose calculation of realistic treatment plans.

We have recently made a tremendous progress towards the further improvement of our gDPM code with emphases on the following two aspects. First, while focusing on algorithm optimization for GPU architecture, we also maintain same particle transport physics as in the original DPM so as to ensure simulation accuracy. In this v2.0 version, it is found that the efficiency is 69.1 ~ 87.2 times higher than the original DPM package on CPU and the CPU and the GPU results are in well agreement. Second, our development also emphasizes on clinical practicality of our code by integrating various key components such as generating particles according to planned fluence maps. To our knowledge, gDPM is the first GPU based MC packages that enables the dose calculation for real treatment plans. It is capable of calculating dose in an IMRT plan or a VMAT plan in a sub-minute time scale.

The roadmap of this paper is as follows. In Section 2, we will briefly describe DPM physics, algorithm structure, and some key issues in our implementation. Section 3 presents experimental results of our dose calculation in both heterogeneous phantoms and realistic IMRT and VMAT plans. Finally, we conclude our paper in Section 4 and present some discussions.

## 2. Methods and Materials

*2.1 DPM Physics*

The original sequential DPM MC code was previously developed for fast dose calculations in radiotherapy treatment planning (Sempau *et al.*, 2000). It targets at simulating coupled photon-electron transport with a set of approximations valid for the energy range considered in radiotherapy. Specifically, the photon transport is handled by using the Woodcock tracking method which greatly increases the simulation efficiency of the boundary tracking process (Woodcock *et al.*, 1965). As for the electron transport, DPM implements a condensed history technique. Step-by-step simulation is used for inelastic collisions and bremsstrahlung emission involving energy losses above certain cutoffs. It also employs new transport mechanics and electron multiple scattering distribution functions to allow long transport steps. These mechanisms enable the electron cross a few heterogeneity boundaries in one step and hence increase the simulation efficiency. The continuous slowing down approximation is employed for energy losses below some preset energy thresholds. Positron transport is treated as electron and two photons are created at the end of the positron path to account for the annihilation process. The accuracy of DPM has been demonstrated to be within ±2% for both clinical photon and electron beams (Chetty *et al.*, 2002; Chetty *et al.*, 2003). In our gDPM code, we will maintain all the physics unchanged, while seeking for efficiency boost using various computational techniques.

*2.2 CUDA implementation*

A GPU typically consists of a large number of scalar processor units. Though the clock speed for each processor is lower than a typical CPU, the overall computational power is much higher due to the large amount of processors available on GPU. Our gDPM v2.0 is coded under the Compute Unified Device Architecture (CUDA) platform developed by NVIDIA (NVIDIA, 2010b), which enables us to extend the C language to program an NVIDIA GPU.

*2.2.1 Simulation process*

The main structure of our code is shown in the left panel of Fig. 1. Once the simulation starts, the code is initialized with all the necessary data including the voxelized geometry, material properties, and all cross section data. Random number seeds are also initialized in this step. All of these data are transferred to GPU memory during this step. After the initialization stage, simulation is then performed in a batched fashion. We evenly divide the total number of histories into $N_b$ batches, *e.g.* $N_b = 10$. For each batch, we simulate the particle transport and record the dose deposition to each voxel. The details of this step will be discussed later. After the simulations for all batches, statistical analysis is performed to obtain average dose to each voxel and corresponding uncertainties. Finally, the program transfer data from GPU to CPU and output results before exit.

Within each batch, particle transport will be simulated in parallel with each GPU thread responsible for one particle. In this process, divergence between GPU threads will take place mainly due to 1) the different physical transport processes that photons and electrons undergo and 2) the randomness of the transport process for a given particle. Though the thread divergence caused by the second factor is hard to remove, by carefully designing the simulation scheme, we can separate the photon and electron transport and hence partially relieve the thread divergence issue.

As such, we design the following data structure. We first allocate a particle array of length *N* to store all the particles currently being simulated. The particles in this array can be either photons or electrons, but all of same type at any time during the simulation. The size of this particle array should be large enough, so that GPU can fully exploit its parallelization ability, while not too large to fit in the GPU memory. Moreover, since GPU execute simulation in warps, *i.e.* a group of 32 threads run simultaneously on a multiprocessor, it is beneficial to choose $N$ to be a multiple of warp size to avoid of wasting resources. In a typical setup, we choose $N = 65536$. Meanwhile, we use two more arrays as stacks to store any secondary particles generated during the simulation, one of a length $M_p$ for photons and one of a length $M_e$ for electrons. The lengths of them are chosen empirically, *e.g.*, $M_p = M_e = 16N$, to provide large enough space for secondary particles.

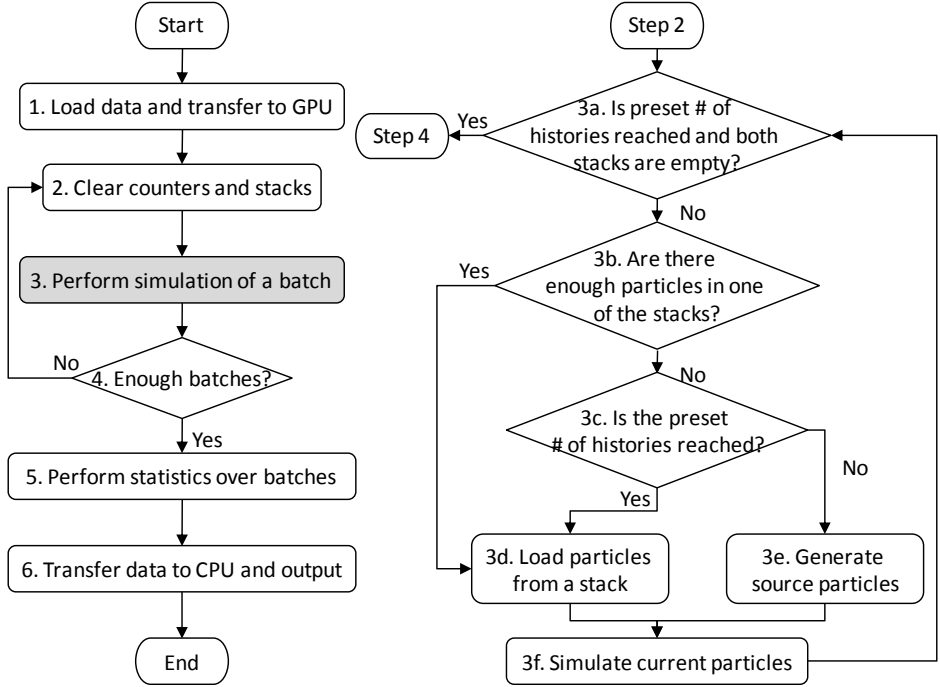

**Figure 1**. The flow chart of our gDPM v2.0 Monte Carlo simulation. Detailed steps of the batch simulation part (Step 3 in the left panel) are shown on the right panel.

Along with this data structure, we design a scheme to carry out the simulation within each batch, such that photon and electron transport are not performed simultaneously. The idea of separating the simulation of electrons and photons was originally proposed by Hissoiny *et al.* (2011) and was found to be effective to speed up MC simulation on GPU. This scheme is illustrated in the right panel of Fig. 1. Specifically, when the program enters Step 3, all the counters and stacks are empty, as well as the particle array. The exit criterion of this Step 3 is that a preset number of particle histories for this batch, *i. e.* the total number of particle histories divided by the number of batches, have already been simulated and both the electron and the photon stacks are empty. Step 3a justifies this criterion and the program exits the current batch, if the criterion is met. Step 3b checks the number of particles in both stacks. If there are more than $N$ particles left in either of them, $N$ particles will be loaded from the corresponding stack to the particle array (Step 3d). If neither stack has enough particles, Step 3c will guide the program to either loading the rest of particles from the stack (Step 3d) in the case all source particle histories have been simulated, or generating new source particles in the particle array (Step 3e) otherwise. Up to now, the particle array is filled with a set of particles, all of same type. Depending on the particle type, a GPU kernel for the electron transport or for the photon transport will be launched in Step 3f. Such a kernel will be executed on GPU in parallel and each GPU thread simulates one particle in the array, till it exits the phantom geometry or is absorbed. In this process, dose deposition to each voxel is recorded and any secondary particles created are put into the corresponding stack. After Step 3f, the program loops back to Step 3a to check if further simulation is needed.

*2.2.2 Other modifications*

Besides turning the sequential simulation into parallel computation utilizing the scheme discussed in section 2.2.1, there are three other modifications we have made in the gDPM v2.0 compared with the original sequential DPM. First, single-precision floating point data type is used throughout our implementation to represent rational numbers instead of double-precision as in the original sequential DPM code. This is due to the fact that single-precision is sufficient for Monte Carlo dose calculation as demonstrated later in this paper.

Second, we have also modified the pseudo-random number generator. In gDPM v2.0, the pseudo-random numbers are generated using the random number generator library CURAND (NVIDIA, 2010a) provided by NVIDIA. This library provides a light-weighted GPU function that produces simple and efficient generation of high-quality pseudo-random numbers using XORWOW algorithm (Marsaglia, 2003), a member of the xor-shift family of pseudorandom number generators. The period of such a generator is about $2^{192}$ and the quality of the random numbers has been tested using the TestU01 "Crush" framework of tests (L'Ecuyer and Simard, 2007).

In addition, we have also modified the interpolation method for the cross section data for better performance. In an MC simulation, an array of a cross section data $\{\sigma_i\}$ at a number of discrete energy values $\{E_i\}$ is first loaded into the memory for each type of interaction of interest. Interpolation of the cross section data is then necessary during the simulation to obtain the cross section value at any arbitrary energy. In the original sequential DPM code, cubic spline interpolation was implemented. As such, interpolation coefficients $\{a_i, b_i, c_i, d_i\}$ are first computed during the program initialization stage for each energy interval $[E_i, E_{i+1}]$, such that the cross section data can be obtained as $\sigma(E) = a_i + b_i E + c_i E^2 + d_i E^3$ for $E \in [E_i, E_{i+1}]$. Though this provides a high accuracy of the interpolation data, each interpolation task requires four times memory read to obtain the four interpolation coefficients and a number of arithmetic operations, which is not efficient considering the slow GPU memory access speed. In the gDPM v2.0, we use linear interpolation instead, which requires only two memory read per interpolation task and less number of arithmetic operations. Moreover, this linear interpolation can be achieved by GPU hardware via the so called texture memory. Since interpolation is very frequently used in the MC simulation, this modification on the interpolation method enhances the overall program efficiency considerably. It is true that the linear interpolation attains lower accuracy on the cross section data. Yet, no loss of accuracy in the final dose has been observed in all of our testing cases.

*2.3 Other components in gDPM v2.0*

In addition to improving computational efficiency, we also focus on clinical practicality of our gDPM package. We have developed a set of necessary components, so that gDPM can be used to compute radiation dose in clinically realistic contexts. First of all, an interface of the gDPM package is built to load clinical treatment plans in the format of DICOM RT. This includes the functions of loading voxelized patient CT data, organ structure information, fluence map of a treatment plan, *etc.*. Second, computations related to linac geometry of a treatment plan are enabled to take into account rotations of linac gantry, multi-leaf collimator (MLC), collimator, and couch. With all the functionalities presented in this section, we are able to load a realistic IMRT or VMAT treatment plan and perform does calculation using the developed gDPM v2.0 package.

*2.3.1 Fluence map*

To simulate IMRT or VMAT treatment plans, generating source particles according to a designed photon fluence map is a key step. Let us group fluence maps from all beam angles together, and divide the whole fluence map from all beam angles into a total number of $N_f$ small beamlets labeled in a certain order by an index $I = 1,2,...N_f$. Note that the index $I$ parameterizes both the beam angle and the location of a beamlet inside an angle. The associated beamlet intensity $f_I$ represents the relative probability that a particle comes from the beamlet $I$. The goal of sampling a photon following this fluence map can be achieved by first sampling a beamlet index $I$ according to the relative probability determined by $f_I$ and then sampling the particle inside this beamlet uniformly. To make this sampling more efficient, let us exclude those beamlets with zero photon fluence from consideration, so that $f_I \neq 0$ for all $I$. A straightforward way of generating source particles according to this fluence map is to use the accumulative probability $P(I) = \sum_{J<I} f_J / \sum_{J<N_f} f_J$. Each time a particle is to be sampled, we can first generate a random number $r$ uniformly distributed in $[0,1]$ and find the beamlet $I$ such that $P(I) \leq r < P(I+1)$. Though this method is conceptually simple, its implementation on GPU is quite inefficient, because the step of finding the beamlet index $I$ satisfying $P(I) \leq r < P(I+1)$ requires some sort of searching algorithms, which involves a large number of memory access to the slow GPU memory.

To ensure high efficiency, we utilize the so called Metropolis sampling algorithm (Hastings, 1970). Key steps of this algorithm are illustrated in Algorithm A1. In this algorithm, the beamlet index generated for the previous one particle $I_{prev}$ is stored, used, and updated each time a new beamlet index is generated.

**Algorithm A1:**

Initialize $I_{prev}$ with an arbitrary beamlet index in $\{1,2,...,N_f\}$.

Do the following steps each time a particle is generated:

1. Generate a trial beamlet $J \in \{1,2,...,N_f\}$ with equal probablity;
2. Generate a random number $r$ uniformly distributed in [0,1];
3. If $r < f_J/f_{I_{prev}}$, set $I = J$; otherwise set $I = I_{prev}$.
4. Generate a particle from the beamlet $I$ uniformly.
5. Set $I_{prev} = I$.

For the initialization step and in the Step 1, the beamlet index $I_{prev}$ and $J$ can be simply chosen from $\{1,2,...,N_f\}$ with equal probability. It has been proven that such an algorithm is able to generate a sequence of beamlet indices following the distribution governed by $f_I$, given that this sequence is long enough. Note that each time a new beamlet index is selected, searching through the fluence map is not involved, which ensures the computational efficiency by avoiding a large number of GPU memory access. In practice, since we are performing parallel computation, each GPU thread is initialized with its own $I_{prev}$ generated by a CPU random number at the initialization stage.

To demonstrate the convergence of this algorithm, we record the particles generated at one beam angle in a typical IMRT plan. The desired fluence is shown in Fig. 2(a), while the photon fluence with $10^7$ particles generated according to the algorithm A1 is depicted in Fig. 2(b). These two fluence maps are visually very similar. To quantify this similarity, we compute the error $e = \|p_f - p_f^*\|_2$, where $p_f$ and $p_f^*$ are vectors composed of the probability at each beamlet for the generated fluence map and the ground truth, respectively. As shown in Fig. 2(c), this error monotonically decreases quickly, as the particle number increases. Considering the number of simulated particles is of order $10^8$ in those realistic treatment plans presented in this paper, the generated fluence is in well agreement with the given fluence map in treatment plans.

*2.3.2 Energy spectrum*

Photons coming from a real linac head are not monoenergetic. Therefore, the source particle energy has to be generated according to an energy spectrum for accurate dose calculation. A straightforward way of taking this spectrum into simulation is to randomly

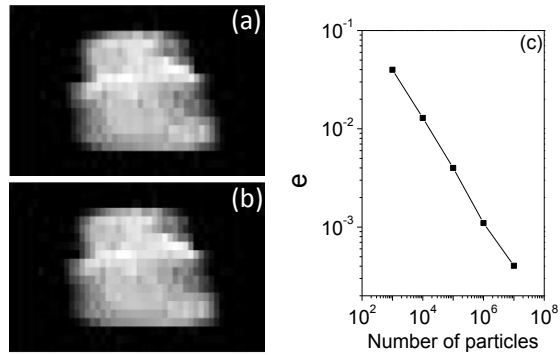

**Figure 2**. A typical fluence map at one beam angle of an IMPT plan is shown in (a). The generated fluence map with $10^7$ particles is shown in (b). (c) is the dependence of the error *e* on the particle number simulated.

9                                                                                                              X. Jia *et al.*assign the energy for each source photon accordingly. Yet, a large number of photons are simulated simultaneously on GPU and the computation time among them varies due to their different energies. As a consequence, those photons with short simulation time will have to wait for those with long simulation time, which reduces the overall computational efficiency. To resolve this issue, we evenly divide the entire energy spectrum into a set of intervals. The total number of photons to be simulated in each batch is first distributed to each energy interval according to the spectrum. Simulation is then performed for each interval sequentially with the particle energy evenly distributed inside the interval. This strategy ensures that, at any moment of the simulation, all GPU threads are dealing with particles of similar energies, removing the possibility of losing efficiency due to the variance of simulation time between GPU threads handling photos of different initial energies.

*2.3.3 Detailed source modeling*

Accurate source modeling of a linac is not a trivial problem by itself and is beyond the scope of this paper. In gDPM v2.0, we do not provide specific source models of any particular type of linac. Instead, we leave the interface of source particle generation part open and users can supplement their own functions to generate source particles according to their own source model or simply by using a phase space file. For a testing purpose, all the photon cases studied in this paper use a point source with a realistic energy spectrum.

**3. Results**

In this section, we provide the results in various cases for testing our gDPM v2.0 package. The purposes of presenting these results are twofold. First, though we did not alter DPM physics, the various techniques employed in gDPM, such as linear interpolation on cross section data, may impact the simulation accuracy. By comparing the simulation results obtained from the sequential DPM and our gDPM v2.0 in phantoms with different materials, we will show that the implementation of our gDPM does not degrade the computational accuracy. Second, the computation time is recorded and compared with that of the original sequential DPM code. This will clearly demonstrate the gain of computational efficiency achieved through the various techniques we employed and the powerful GPU we used. In addition, dose calculation for an IMRT case and two VMAT cases will also be conducted to demonstrate the feasibility of using gDPM for fast MC dose calculation in realistic clinical contexts.

As for the hardware used in this section, the GPU results are obtained on an NVIDIA Tesla C2050 card. Such a GPU card is manufactured specifically for the purpose of scientific computing. It has a total number of 448 processor cores (grouped into 14 multiprocessors with 32 cores each), each with a clock speed of 1.15 GHz. The card is equipped with 3 GB GDDR5 memory shared by all processor cores. It supports error correction codes to protect data from random errors occurred in data transfer and manipulation, ensuring computing accuracy and reliability. As for the CPU on which the original DPM code is executed, it is equipped with a 2.26 GHz Intel Xeon E5520 Nehalem processor and 4GB memory.

*3.1 Phantom studies*

We first study the performance of our gDPM v2.0 on two phantoms with slab geometries. The dimension of both of them are $30.5 \times 30.5 \times 30.0$ cm$^3$ and the voxel size is set to be $0.5 \times 0.5 \times 0.2$ cm$^3$. The first phantom consists of three layers along the *z* direction, namely 5 cm water, 5 cm bone, and 20 cm water. The geometry of the second phantom is same as the first one except that the bone slab is replaced by a lung slab. In all testing cases, we place a point source at SSD = 90 cm, and the beam impinges normally to the phantom on its *x-o-y* plane. Field size is set to be $10 \times 10$ cm$^2$ at the isocenter with SAD = 100 cm. The electron beam is mono-energetic with it energy set to be 20 MeV, while the photon beam has a realistic 6 MV energy spectrum. The absorption energies are 200 keV for electrons and 50 keV for photons. For the cases with an electron source, a total number of $2.5 \times 10^6$ particle histories are simulated, which are evenly divided into 10 batches for the purpose of calculating statistical uncertainties. As for the cases with a photon source, $2.5 \times 10^8$ particle histories are chosen and same number of batches is used.

In Fig. 3 and Fig. 4, the left columns are the depth dose curves along the beam's central axis, while the right columns correspond to the lateral dose profiles taken at $z = 2.5, 7.5,$ and $12.5$ cm. The error bars represent the level of two standard deviations of the results. The error bars corresponding to the CPU results are not drawn for the purpose of clarity, which are of similar sizes to those for the GPU results. These figures visually demonstrate good agreements, within statistical uncertainty, between the CPU and the GPU results due to the unaltered physics in our gDPM v2.0 code. To quantify the precision of our simulation, we calculate the uncertainty at each voxel $\sigma$ normalized by the maximum dose $D_{max}$. We further average the relative uncertainty $\sigma/D_{max}$ over the high dose region where the local dose $D$ exceeds half of the maximum value $D_{max}$ in the entire phantom. The quantity $\overline{\sigma/D_{max}}$ indicates the simulation precision in the high dose region. In all four testing cases, we have simulated enough number of particle histories, so that $\overline{\sigma/D_{max}}$ is found to be less than 1% for both the CPU and the GPU results, as indicated in Table 1.

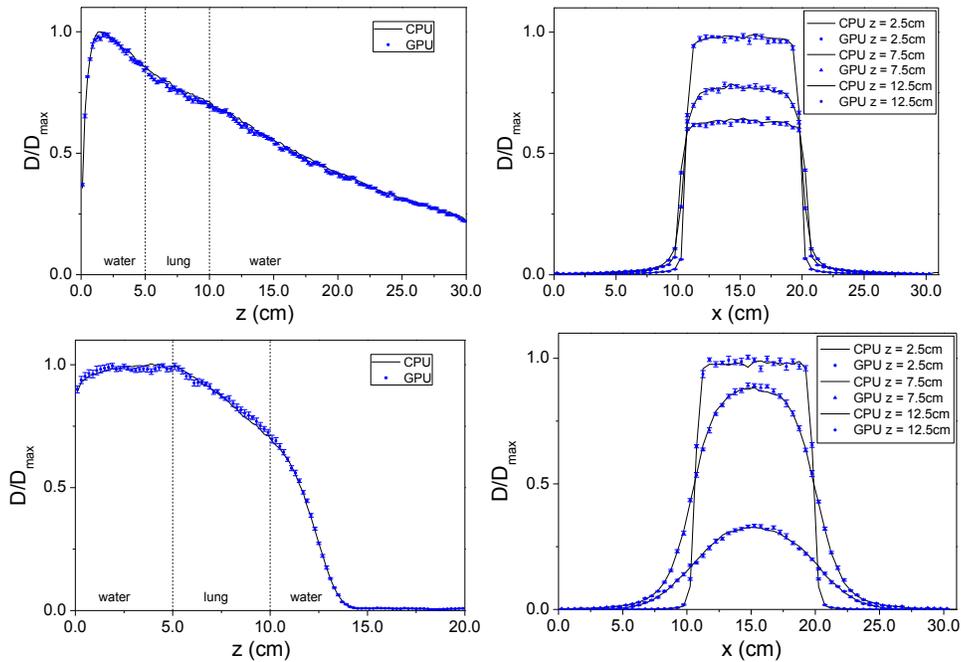

**Figure 3**. Depth-dose curves (left column) and lateral dose profiles at different depths (right column) of a $10 \times 10$ cm$^2$ photon beam (top row) and an electron beam (bottom row) at SSD = 90cm impinging on a water-lung-water phantom.

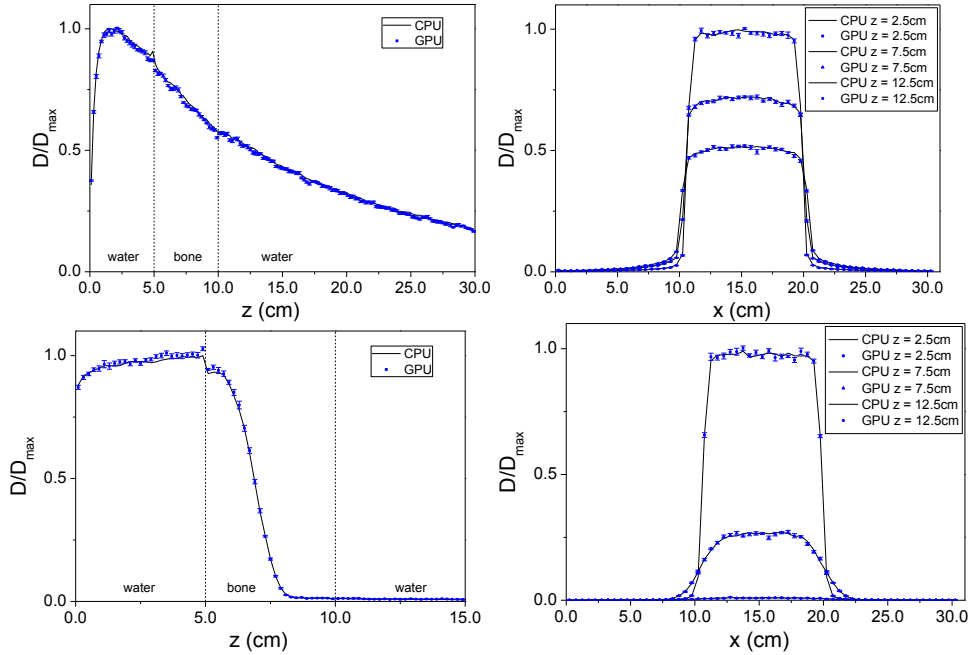

**Figure 4**. Depth-dose curves (left column) and lateral dose profiles at different depths (right column) of a $10 \times 10 \text{cm}^2$ photon beam (top row) and an electron beam (bottom row) at $SSD = 90 \text{cm}$ impinging on a water-bone-water phantom.

We have also quantified the agreement between the CPU results and the GPU results using a statistical two-tailed t-test. The dose value at each voxel is a statistical quantity, which fluctuates each time an MC simulation is performed. The t-test justifies whether the dose values at a voxel obtained from a CPU simulation and a GPU simulation agree with each other under a certain statistical significance level. In this test, the null hypothesis is that there is no difference between the CPU and the GPU results at a given voxel. In practice, we first compute the $t$ value as $t = (D_{CPU} - D_{GPU})/\sqrt{\sigma_{CPU}^2 + \sigma_{GPU}^2}$ at each voxel, where $\sigma_{CPU}$ and $\sigma_{GPU}$ are uncertainties corresponding to the dose $D_{CPU}$ and $D_{GPU}$ at the voxel, respectively. By comparing this $t$ value with a threshold corresponding to a significance level of $\alpha = 0.05$, the hypothesis is accepted or rejected. If the hypothesis is not rejected at a voxel, it is implied that, with 90% confidence, the dose difference at that voxel between the CPU and the GPU simulations is not statistically significant. To quantify the overall agreement between the two dose distributions in high dose region, we finally compute the passing rate of this test $P_{high}$ as the ratio of the number of voxels where the hypothesis is not rejected over the total number of voxels in the high dose region. Similarly, the passing rate $P_{all}$ is computed over the entire phantom region to characterize the overall agreement. As shown in Table 1, it is observed that $P_{high} > 96\%$ and $P_{high} > 97\%$ depending on the cases studied. These high passing rates clearly imply well agreements between the CPU and the GPU results.

**Table 1**. Average relative uncertainty ($\overline{\sigma/D_{\max}}$) and t-test passing rate in high dose region ($P_{high}$) and in entire phantom region ($P_{all}$) for four different test cases.

| Source type | # of Histories | Phantom | $\overline{\sigma/D_{\max}}$ CPU (%) | $\overline{\sigma/D_{\max}}$ GPU (%) | $P_{high}$ (%) | $P_{all}$ (%) |
|---|---|---|---|---|---|---|
| 20MeV Electron | $2.5 \times 10^6$ | water-lung-water | 0.99 | 0.98 | 99.9 | 99.9 |
| 20MeV Electron | $2.5 \times 10^6$ | water-bone-water | 0.98 | 0.99 | 100.0 | 99.8 |
| 6MV Photon | $2.5 \times 10^8$ | water-lung-water | 0.71 | 0.72 | 98.5 | 97.7 |
| 6MV Photon | $2.5 \times 10^8$ | water-bone-water | 0.64 | 0.64 | 96.9 | 97.5 |

Table 2 depicts the computation time for these testing cases. $T_{CPU}$ stands for the execution time of the CPU implementation, while $T_{GPU}$ is that of the GPU implementation including data transfer time between CPU and GPU. Speed-up factors of about 69.1 ~ 87.2 have been observed for the GPU calculation compared to the CPU simulation.

*3.2 Realistic patient cases*

To further demonstrate the feasibility of using gDPM v2.0 for dose calculations in realistic treatment plans, we have also loaded one IMRT plan and two VMAT plans generated from Varian Eclipse treatment planning system (Varian Medical Systems, Inc., Palo Alto, CA, USA). The IMRT plan consists of 8 non-coplanar fields for a head-and-neck (HN) cancer patient, while one of the VMAT plans is for a HN cancer case and the other is for a prostate cancer case, both with two arcs. For the purpose of illustrating the principle, a simplified source model with a 6MV point photon source and a realistic energy spectrum is used for these cases. The calculated dose distribution is displayed in Fig. 5 and the simulation uncertainty and computation time are listed in Table 3. With a total number of $2.5 \times 10^8$ source photons in each case, the relative uncertainty is controlled to be under 1% in the high dose region ($D > 0.5\, D_{\max}$). As for the computation time, it takes 36.1 sec for the dose calculation in the IMRT case and 36.7 sec and 39.6 sec for the two VMAT cases, respectively. Compared with the simulation time for the phantom cases, this time is prolonged, mainly due to the source particle generation from fluence maps.

**Table 2**. Computation time on CPU ($T_{CPU}$), that on GPU ($T_{GPU}$), and the speed up factor ($T_{CPU}/T_{GPU}$) for four different test cases.

| Source type | # of Histories | Phantom | $T_{CPU}$ (s) | $T_{GPU}$ (s) | $T_{CPU}/T_{GPU}$ |
|---|---|---|---|---|---|
| 20MeV Electron | $2.5 \times 10^6$ | water-lung-water | 117.5 | 1.70 | 69.1 |
| 20MeV Electron | $2.5 \times 10^6$ | water-bone-water | 127.0 | 1.65 | 77.0 |
| 6MV Photon | $2.5 \times 10^8$ | water-lung-water | 1403.7 | 16.1 | 87.2 |
| 6MV Photon | $2.5 \times 10^8$ | water-bone-water | 1741.0 | 20.5 | 84.9 |

## 4. Discussion and Conclusions

In this paper, we reported our recent development of a GPU based MC dose calculation package, gDPM v2.0. The code is specifically tailored for the GPU architecture to achieve high computation efficiency. Due to the identical particle transport physics to that in the original DPM code, the accuracy of the gDPM package is not degraded. Simulations in various phantom cases indicate that results from CPU and GPU are in well agreement for both the electron and the photon sources. In particular, statistical t-tests are performed and the dose differences between the CPU and the GPU results are found not statistically significant in over 96% of the high dose region and over 97% of the entire phantom region. With a powerful yet affordable NVIDIA Tesla C2050 GPU, speedup factors of 69.1 ~ 87.2 have been observed against a 2.27GHz Intel Xeon CPU processor. The development of gDPM v2.0 package also focuses on its clinical practicality. With a set of necessary components developed for dose calculation in realistic clinical plans, IMRT or a VMAT plan dose calculation using MC simulation can be achieved in 36.1~39.6 sec with a single GPU.

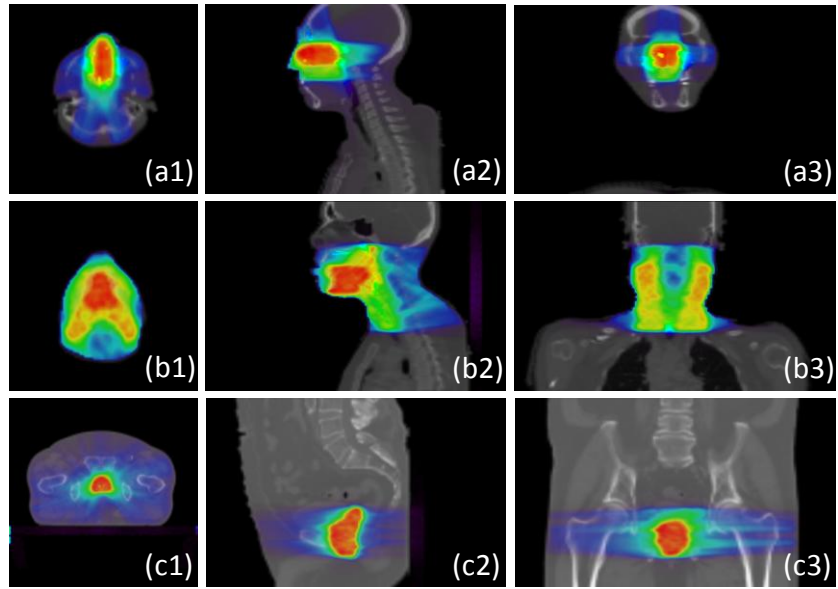

**Figure 5.** Dose calculation results for an 8 noncoplanar beam HN IMRT plan (a1) ~ (a3), a 2-arc HN VMAT plan (b1) ~ (b3), and a 2-arc prostate VMAT plan (c1) ~ (c2) using the gDPM v2.0 package.

**Table 3**. Average relative uncertainty ($\overline{\sigma/D_{\max}}$) and GPU computation time ($T_{GPU}$) for the dose calculation in three realistic treatment plans.

| Case | # of Histories | Size | Resolution (cm$^3$) | $\overline{\sigma/D_{\max}}$ (%) | $T_{GPU}$ (s) |
|---|---|---|---|---|---|
| IMRT HN plan | 2.5×10$^8$ | 128 × 128 × 141 | 0.39 × 0.39 × 0.25 | 0.57 | 36.1 |
| VMAT HN plan | 2.5×10$^8$ | 128 × 128 × 120 | 0.39 × 0.39 × 0.25 | 0.98 | 36.7 |
| VMAT prostate plan | 2.5×10$^8$ | 128 × 128 × 131 | 0.39 × 0.39 × 0.25 | 0.74 | 39.6 |

At the end of this paper, we would like to discuss a few questions that might be of interest to readers. First, is it possible to further improve the computation efficiency? Though considerable speed up factors have been achieved in gDPM v2.0 and it becomes possible to perform MC dose calculation in a sub-minute time scale for realistic clinical treatment plans, it would make the GPU-based MC dose calculation more clinically attractive, if the computation time could be further shortened. In fact, there are a number of ways that could potentially further increase the computational efficiency. (1) If a multi-GPU platform is available, all the particle histories simulated can be simply distributed among all the GPUs, which then execute simultaneously without interfering with each other. Only at the end of the computation will the dose distribution be collected from all the GPUs. Due to the negligible overhead in this process, it is expected that a roughly linear scalability of the computation efficiency can be achieved with respect to the number of GPUs. In a recently work, it has been reported that this linear scalability holds at least on a dual-GPU system (Hissoiny *et al.*, 2011). We have also tested the multi-GPU performance of our gDPM code on a 4-GPU platform. In our tests, a bash shell script submits simulation jobs to all GPUs simultaneously and another small program is then launched by the script after all simulations are completed to accumulate simulation results. Speed up factors of 3.98~3.99 compared to a single GPU have been observed amount various test cases. These observations clearly demonstrate the simplicity yet feasibility of achieving a further efficiency boost utilizing a multi-GPU platform. In particular, this 4-GPU platform will bring MC dose calculation time for realistic plans under 10 seconds. Currently, a multi-GPU version of our gDPM package using Massage Passing Interface is under development. (2) Since no variance reduction technique is employed in gDPM v2.0, the convergence rate of the calculated dose suffers a lot from the stochastic nature of the particle transport process. With the integration of variance reduction techniques such as particle splitting and track repeating, it is expected that further improvement of the efficiency can be achieved.

Second, how is the performance of the gDPM v2.0 compared with other similar work? Recently, another work on GPU-based MC dose calculation, GPUMCD, has been reported by Hissoiny *et. al.*(2011), where the simulation scheme of separating photon and electron transports was invented. To test our gDPM v2.0 against GPUMCD, we run our code in a homogeneous water phantom with a resolution of $64 \times 64 \times 64$ voxels and the voxel size is $0.5 \times 0.5 \times 0.5$ cm$^3$, same as in the first testing case reported in Hissoiny *et al.*. A mono-energetic (15 MeV) mono-directional photon source or an electron source normally impinges on the phantom. Photon generations in the positron annihilation process are switched off as in the GPUMCD for a fair comparison. 1 million and 4 million particles are simulated for the electron source case and the photon source case, respectively. The absolute running time is 0.47 sec for the electron source case and 0.90 sec for the photon source case on a Tesla C2050 GPU card. Compared to the running time of 0.12 sec and 0.27 sec for GPUMCD in the two cases on a GTX 480 card, our code is ~4 times slower. This speed difference can be first ascribed to the different particle transport physics employed in two packages. Our gDPM package does not change the complicated yet accurate DPM physics, while GPUMCD combines various particle transport physics described in general-purpose MC packages and implements them for GPU calculation. The different cut-off energies used in the two packages also change the simulation efficiency. Moreover, part of the efficiency difference comes from different hardware. Since C2050 is manufactured by NVIDIA dedicated for scientific computing, it scarifies its efficiency for accuracy and reliability to some extent. For instance, the GPU core speed, memory speed, and memory bandwidth of GTX 480 are 21.7%, 23.2%, and 22.9% higher than those for the C2050, respectively (Wikipedia).

Yet, detailed comparisons on the computational efficiency between GPUMCD and gDPM are not of critical importance, as both of them have achieved high enough efficiency for clinical applications. Though gDPM v2.0 is not currently the fastest GPU-based MC simulation package, its development balances the speed and accuracy. In particular, despite the various numerical techniques utilized in gDPM, it has been demonstrated that similar accuracy level to the original DPM code can be achieved in gDPM, which has been previously shown to be within $\pm 2\%$ for both clinical photon and electron beams. Moreover, to our knowledge, gDPM is the first GPU based MC packages that enables the dose calculation of realistic treatment plans. Clinical implementation of gDPM will offer high accuracy MC dose calculation in cancer radiotherapy in a sub-minute time scale.

**Acknowledgements**

This work is supported in part by the University of California Lab Fees Research Program. We would like to thank NVIDIA for providing GPU cards for this project. We would also like to acknowledge Sami Hissoiny from Ecole Polytechnique de Montréal and Dr. Joseph Sempau from Universitat Politècnica de Catalunya for fruitful discussions.